\documentclass[twocolumn,times]{aastex6}

\usepackage[english]{babel}
\usepackage[utf8x]{inputenc}
\usepackage[T1]{fontenc}

\usepackage[a4paper,top=3cm,bottom=2cm,left=3cm,right=3cm,marginparwidth=1.75cm]{geometry}

\usepackage{ulem}
\usepackage{graphicx}
\usepackage{amsmath}
\usepackage{epsfig} 
\usepackage{enumitem}
\usepackage[colorinlistoftodos]{todonotes}

\begin{document}

\title{The GROWTH Marshal: A Dynamic Science Portal for Time-Domain Astronomy}

\author{M. M. Kasliwal\altaffilmark{1}, C. Cannella\altaffilmark{1}, A. Bagdasaryan\altaffilmark{1}, T. Hung\altaffilmark{2}, U. Feindt\altaffilmark{3}, L. P. Singer\altaffilmark{4}, M. Coughlin\altaffilmark{1}, C. Fremling\altaffilmark{1}, R. Walters\altaffilmark{1}, D. Duev\altaffilmark{1}, R. Itoh\altaffilmark{5}, R. M. Quimby\altaffilmark{6}}

\altaffiltext{1}{Division of Physics, Mathematics and Astronomy, California Institute of Technology, Pasadena, CA 91125, USA}
\altaffiltext{2}{Department of Astronomy, University of Maryland, College Park, MD 20742, USA}
\altaffiltext{3}{The Oskar Klein Centre, Department of Physics, Stockholm University, AlbaNova, SE-106 91 Stockholm, Sweden}
\altaffiltext{4}{Astroparticle Physics Laboratory, NASA Goddard Space Flight Center, Mail Code 661, Greenbelt, MD 20771, USA}
\altaffiltext{5}{Department of Physics, School of Science, Tokyo Institute of Technology, 2-12-1 Ohokayama, Meguro, Tokyo 152-8551, Japan}
\altaffiltext{6}{Department of Astronomy / Mount Laguna Observatory, San Diego State University, 5500 Campanile Drive, San Diego, CA, 92812-1221}
\altaffiltext{7}{Kavli Institute for the Physics and Mathematics of the Universe (WPI), The University of Tokyo Institutes for Advanced Study, The University of Tokyo, Kashiwa, Chiba 277-8583, Japan}

\begin{abstract}
We describe a dynamic science portal called the GROWTH Marshal that allows time-domain astronomers to define science programs, program filters to save sources from different discovery streams, co-ordinate follow-up with various robotic or classical telescopes, analyze the panchromatic follow-up data and generate summary tables for publication. The GROWTH marshal currently serves 137 scientists, 38 science programs and 67 telescopes. Every night, in real-time, several science programs apply various customized filters to the 10$^{5}$ nightly alerts from the Zwicky Transient Facility. Here, we describe the schematic and explain the functionality of the various components of this international collaborative platform. 

\end{abstract}

\section{Introduction}

New time-domain surveys, with their immense survey mapping speeds, are taking a leap by an order of magnitude in the sheer volume of their discovery streams. Ongoing optical time-domain surveys optimized for various cadences and depths include ASAS-SN \citep{asassn}, ATLAS \citep{atlas}, DLT40 \citep{dlt40}, Evryscope \citep{evryscope}, MASTER \citep{master}, PanSTARRS \citep{panstarrs} and the Zwicky Transient Facility (ZTF; Bellm et al. in prep, Graham et al. in prep). For example, ZTF is already generating alerts of the order of 10$^{5}$ events per night and soon, starting 2022, the Large Synoptic Survey Facility (LSST; \citealt{lsst}) will generate alerts of the order of 10$^{6}$ events per night. An alert is defined as any candidate astronomical source that has changed in flux density (or is at a new spatial position) relative to an archival reference image.

A major challenge that faces the astronomical community is how to efficiently work with such large datasets to identify well-defined samples of sources of interest and obtain the necessary follow-up data in a timely manner. Especially with worldwide collaborations involving multiple follow-up telescopes, organization, co-ordination and communication are key to an effective, productive scientific collaboration. Every system undertaking systematic follow-up of transients is putting together tools to facilitate this process, e.g. the PESSTO collaboration \citep{pessto} and Las Cumbres Observatory \citep{lco}.   

Here, we present a dynamic web science portal that addresses this challenge dubbed the GROWTH Marshal. The word marshal signifies that this portal is designed to marshal transient candidates to follow-up telescopes (or vice versa). GROWTH (Global Relay of Observatories Watching Transients Happen) is a worldwide network of 16 institutions committed to time-domain astronomy and it is the name of the NSF PIRE (National Science Foundation Partnership in International Research and Education) project that developed this portal. The codebase is written by multiple students and postdocs and builds on the legacy of code developed for the Palomar Transient Factory (PTF; \citealt{ptf}) project. The codebase is written entirely in Python in modular form and version controlled. All data is stored in a postgreSQL database.   

\begin{figure*}[!hbt]
\centering
\includegraphics[width=0.9\textheight,angle=90]{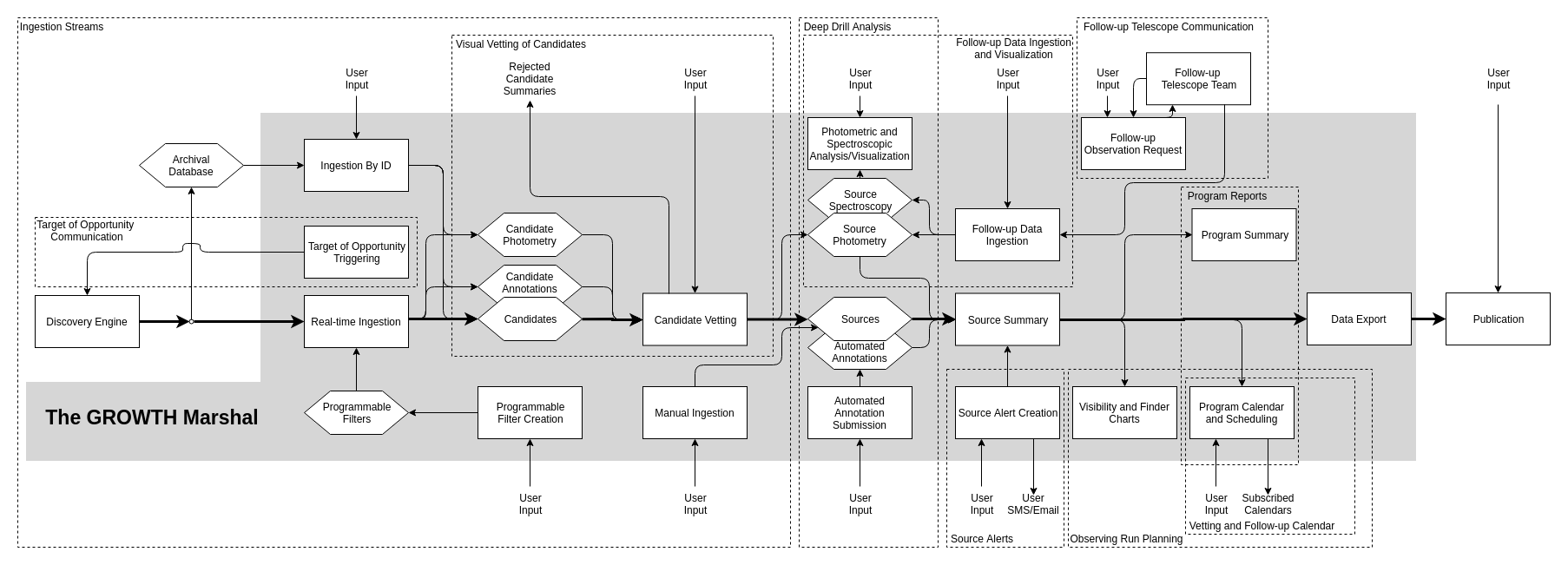}
\caption{\label{fig:schematic} Schematic representing the various components of the GROWTH Marshal Science Portal.}
\end{figure*}

Currently, the GROWTH marshal is being used by 137 scientists with 38 science programs and 67 telescopes worldwide. Each science program has a well-defined sample-selection criterion to filter candidates from various discovery streams. Every night, members of a science program monitor which sources pass their filter. Based on the available information, users regularly marshal transient candidates to various follow-up telescopes and analyze the resulting follow-up data. A succinct summary of discovery stream data, follow-up data and analysis results is presented on the science program reports page. This summary is machine-readable to ease regular monitoring as well as ease generation of tables for journal publication. The key service of the GROWTH marshal is to assist a science collaboration with the steps between receiving a firehose of alerts from a discovery engine to publication of the science results. Figure~\ref{fig:schematic} shows a schematic summarizing the various components of the GROWTH marshal. 

\begin{figure*}[hbt!]
\centering
\includegraphics[width=0.95\textwidth]{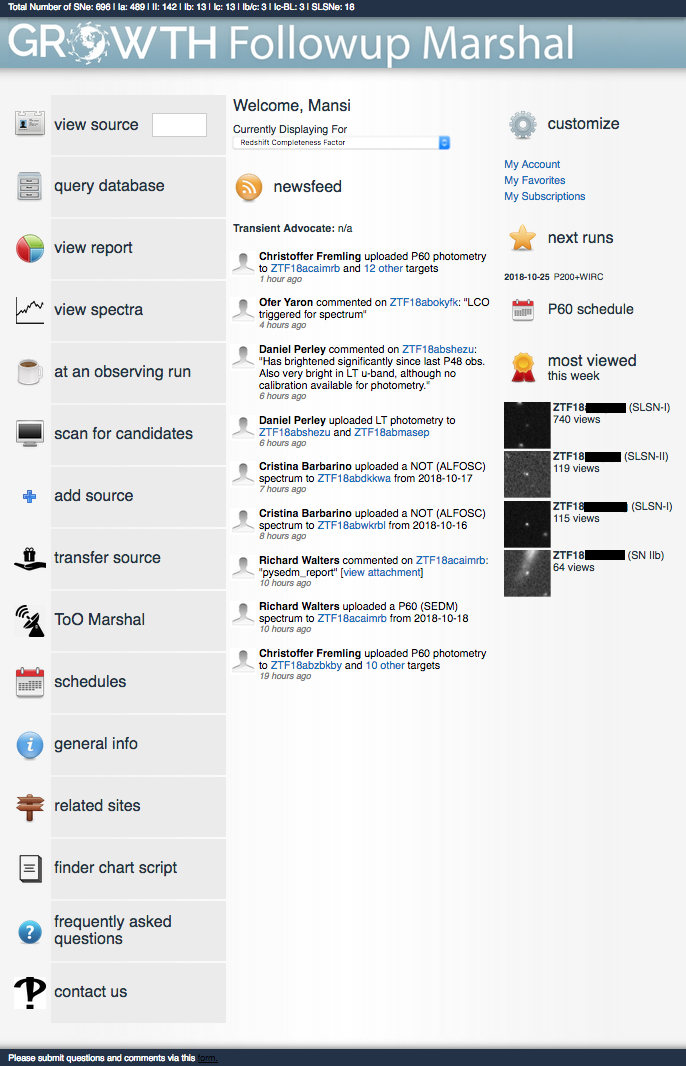}
\caption{\label{fig:homepage}The GROWTH Marshal Homepage.}
\end{figure*}

The GROWTH Marshal's home page (Figure~\ref{fig:homepage}) acts as a unified, central hub allowing users to easily access various pages.  The home page displays the most viewed sources of the week, letting users quickly navigate to the most exciting and interesting new sources. The home page features a newsfeed updating the user on the latest comments, classifications, and observation assignments for sources in their science programs. Each science program also has a calendar to organize candidate vetting and/or follow-up observing runs. The backend is a Google Calendar API such that users may subscribe to these calendars with other calendar applications to view them along with other calendars. This home page may also be limited to particular science programs, allowing users to quickly see what is happening in any of their science programs, not just those programs that happen to be the most active.    

The home page provides easy access links for the various facets of a user's science program. It has links to ingestion streams (\S~\ref{sec:ingestion}), visual vetting of filtered candidates (\S~\ref{sec:vetting}), deep-drill analysis of selected sources (\S~\ref{sec:sources}), communicating with robotic telescopes (\S~\ref{sec:robotic}), planning tools for observing runs (\S~\ref{sec:observingrun}), visualizing follow-up data (\S~\ref{sec:followup}), viewing a summary as a science program report (\S~\ref{sec:report}) and triggering Target of Opportunity programs (\S~\ref{sec:too}). 

\section{Ingestion Streams}
\label{sec:ingestion}

There are three avenues by which new sources may be saved on the GROWTH Marshal: programmable real-time filtering, archival retrieval, and manual entry. Each of the current methods of saving sources to the GROWTH Marshal is intended to allow our users to pick out sources they find relevant to their science programs and to avoid missing any interesting sources among a vast number of irrelevant candidates.  

First, throughout a night of observation, the GROWTH Marshal promptly receives a stream of Avro packets (Patterson et al. in prep) from the Zwicky Transient Facility describing properties of recently detected candidate sources, including a rolling 30-day history of detections and non-detections of that candidate.    
Users of the GROWTH Marshal write programmable filters in a purpose-built domain-specific language to automatically accept or reject detected candidates in real-time as Avro packets are received by the Marshal. A user's programmable filter defines a function to be applied to each incoming candidate's Avro packet that returns a boolean value representing whether or not the source passes a science program's selection criterion.  The programmable filter language is intentionally simplified to allow the GROWTH Marshal to safely attempt to execute any written filter and to ensure that all candidates can be filtered for every programmable filter in near real time.  Any candidate that passes a user's programmable filter is presented to the user for manual candidate vetting prior to saving it as a source on the Marshal. The programmable filters do not perform any image-level operations. 

Second, for users that may wish to perform more extensive offline calculations to select sources relevant to their science programs, we allow archival retrieval.
Each detected candidate's Avro packet is fully archived in an external database once the packet is broadcast.
The GROWTH Marshal is able to retrieve archived information about a candidate using the candidate's Avro packet id and thus save particular candidates as specified by the user.  All candidates saved by their Avro packet ids are still presented to the user for manual candidate vetting before being saved as a source on the Marshal.

Third, we recognize that users may be interested in follow-up of sources discovered by surveys other than the Zwicky Transient Facility and using manual discovery streams such as announcements on Astronomer's Telegram or the Gamma-ray Circular Network.  Thus, the GROWTH Marshal allows users to manually add sources to their science programs by providing the source's name and position.  We assume that there is no uncertainty about the user's interest in a manually added source and that the user has already sufficiently vetted such a source.  Therefore, manually added sources skip the candidate vetting step and are immediately saved as sources on the Marshal under a particular science program.

Finally, to further facilitate collaboration, the GROWTH Marshal includes a system that allows manual transfer, sharing and deletion of sources among science programs. This is needed if a source is interesting to several programs simultaneously, or if follow-up deems that a source no longer satisfies the criteria for the science program that saved it. For example, if a core-collapse science program finds that a candidate is a thermonuclear supernova after spectroscopic classification, any member of the core-collapse program may choose to either share or transfer the source from their program. 

\section{Visual Vetting of Candidates}
\label{sec:vetting}

The GROWTH Marshal has a graphical user interface that presents all candidates saved either by passing a user's programmable filter or by packet id to the user for further manual vetting. Each candidate is presented with a triplet of image-cutouts centered on the candidate position in the reference image, science image and difference image. A user's candidates are listed alongside more detailed information for each candidate such as the candidate's photometric evolution.  Users are able to manually save or reject candidates based on the provided information.  Saved candidates are immediately saved as sources on the Marshal.  Rejected candidates are noted along with the reason for rejection, which is then made available to the candidate stream's developers to improve the performance of machine-learning algorithms and other candidate detection apparatus. A candidate that is not selected by one science program is allowed to be selected by another science program. Candidates can be rank-sorted by their machine-learning score (Mahabal et al. in prep). Alternately, if users plan to follow-up candidates in real-time, they can sort the stream of candidates by observation time so that they look at the newest candidates first.  

During development of a filter, we anticipate that science programs may have too large a number of candidates passing a users filter that are impractical to fully review.  Thus, the GROWTH Marshal's programmable filters allow users to save and label values calculated during the application of the filter to incoming candidate Avro packets (Patterson et al. in prep).  These saved values are presented alongside candidates as annotations for use in vetting.  The GROWTH Marshal provides the ability to sort candidates by the value of annotations with a particular label, allowing users to prioritize vetting candidates that may be the most interesting for their purposes. This allows users to re-tune their filters so that a more manageable number passes their selection criterion.

As the ZTF discovery stream evolves, machine-learning improves and the GROWTH marshal filters are fine-tuned, we anticipate that this manual candidate vetting step will not be needed to separate real astrophysical sources from bogus sources. The key to automate the saving of candidates as sources is selection criterion that yield a sufficiently pure stream of astrophysically real sources that each satisfy a science program's selection criterion. 

\section{Deep-drill Analysis on Selected Sources}
\label{sec:sources}

Once a candidate is saved to a science program, the members perform deep-drill analysis to a source to help prioritize follow-up observations and provide additional clues on the nature of the source (Figure~\ref{fig:viewsource}). All photometry is displayed in an interactive light curve where the user can add/remove filters and zoom in/out. All spectra are shown in an interactive display where the user can overplot line transitions of various elements, sky lines, telluric lines etc. Cross-matches to various archival, multi-wavelength catalogs is done to retrieve relevant information. Users can freely add comments about the nature of the source. Users can do a detailed examination by pulling up all detection image triplets of the source. Each science program has developed different analysis tools customized to their science goals. For example:

\begin{figure*}[ht!]
\centering
\includegraphics[width=1\textwidth]{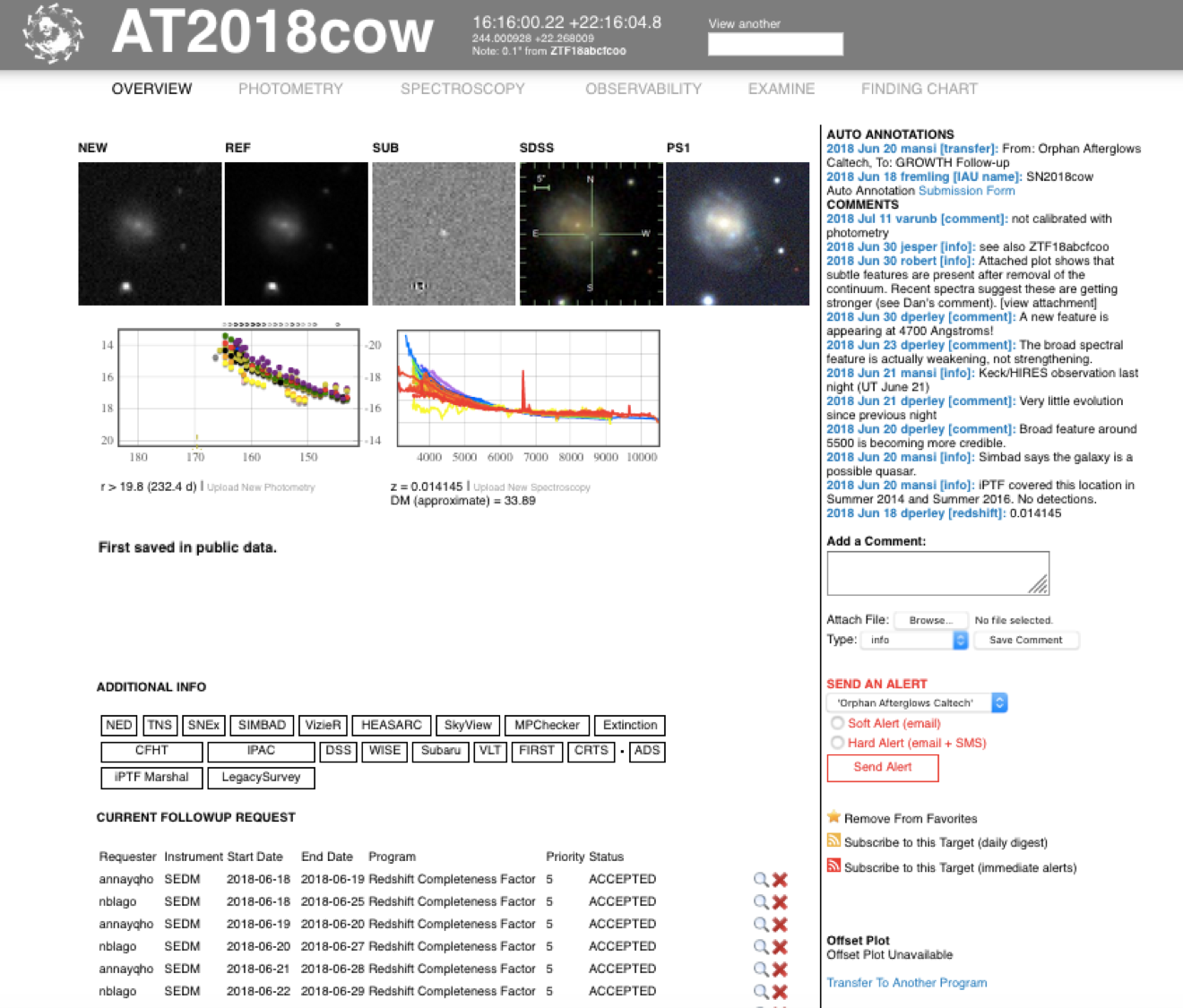}
\caption{\label{fig:viewsource} An example candidate deep-drill page which collates the photometric and spectroscopic follow-up associated with the source, links the candidate to various databases, allows alerts to be sent, annotations to be freely added and follow-up assignments to be made.}
\end{figure*}

\subsection{Light Curve Fitting Tool}

For some classes of transients, users can run a light curve fitter to predict future source brightness and thus plan out which resources, what cadence and what depth should be used for obtaining additional data. Specifically, when using type Ia supernovae as cosmological distance indicators, their peak magnitudes need to be standardized because supernovae that are fading slower are also intrinsically brighter. Similarly supernovae that are bluer at peak are also brighter. To determine this standardization, a SN~Ia template needs to be fit to the lightcurve. The GROWTH marshal has a simple tool that fits the SALT2 template \citep{2007A&A...466...11G, 2014A&A...568A..22B} using the Python package \texttt{sncosmo} \citep{kyle_barbary_2016_168220}. Since this fit requires exact knowledge of the filter transmission curves, the data is limited to photometry from ZTF and a few other telescopes. The lightcurve fit results yield useful information such as the predicted time of peak and magnitude at peak. Additionally comparing the absolute peak magnitude and the ``stretch'' and color parameters, $x_1$ and $c$, respectively, to the expected values for a type Ia supernova can be used to assess whether the transient may be a type Ia supernova or not. Since \texttt{sncosmo} offers easy access to more lightcurve templates, including many for supernova types other than Ia, we are implementing a feature that allows the user to select which template to fit. The quicklook fit results are very valuable for timely followup decisions, even if the transient that the template was fit to is not a type Ia supernova. 

\subsection{Offset Plot}
In extragalactic astronomy, measurement of astrometry has proven to be an effective parameter to identify transient phenomena that are associated with nuclear black holes (e.g. tidal disruption events and AGN variability). Although the offset from the host galaxy centroid is available at the scanning phase, there are usually too few detections to tell whether the offsets are indeed clustered around the host centroid. In light of this, the offset plot is provided as a functionality for the saved candidates on the GROWTH Marshal. The backend Marshal database is designed to archive the parameters denoting the measured offset between the transient and the nearest source in the reference image for every detection. We created a script to display the scatter plot of $\Delta$RA and $\Delta$DEC for each detection by querying the photometry table. On the background, we also mark nearby objects ($<$ 10 arcsec) in the Pan-STARRS1 catalog to help visually associate transients with their host galaxies.

\subsection{Automated Annotations}

\paragraph*{}  An automated annotation for a source on the GROWTH Marshal consists of a name, a datatype, and the annotation's content.  The automated annotation's name is a description of the property described by the annotation, e.g. ``SDSS\_spec\_z''.  Since the automated annotations are intended to act as extensions to the Marshal's database entries and since the Marshal has no ability to decide which automated annotation among many should be regarded as describing the ``true'' value for a particular property, the GROWTH Marshal allows at most one automated annotation with a given name to be present for a particular source.  The automated annotation's datatype, e.g. ``BOOL'', ``FLOAT'', or ``STRING'', is used by the Marshal to determine how to interpret the annotation's content.  The annotation's content consists of a single value consistent with the specified datatype followed by any other relevant information required by the user. When the GROWTH Marshal attempts to interpret an automated annotation, it takes everything in the annotation's content before the first space as the annotations's value to be interpreted in keeping with the annotation's datatype and ignores all of the annotation's content after the first space.  This allows automated annotations to be used by the Marshal for sorting and searching, while also allowing automated annotations to contain further information, like error bars or external hyperlinks, needed for users to properly interpret the automated annotation's value.  

\paragraph*{}  Finally, the GROWTH Marshal allows users to retrieve saved source data from the Marshal and submit new automated annotations via HTTP Post requests.  The Marshal's users are therefore able to determine relevant source properties and submit new automated annotations to the Marshal using scripts on their own machines.  Thus, the submission of automated annotations to the Marshal may be automated by users and customized by each science program.

\subsection{Alerts on sources that require urgent follow-up}
With all the follow-up data and analysis tools at the user's fingertips, a member of a science program may realize that a source requires immediate follow-up (within minutes or hours). Thus, the GROWTH Marshal provides an alert message function to call the attention of all members of a given science program. There are two modes of the alert message: soft alert and hard alert. Soft alerts are sent in the form of emails to members of a science program. Hard alerts are sent as text messages to the mobile numbers of members of a science program. Accidental hard alerts are avoided with a pop-up window to confirm the sending of the alert. The users may also set their preferences to not be disturbed by hard alerts and will receive emails instead. This function improves the inefficiency in traditional email correspondence with a template message for each source. The message specifies the name and coordinates of a source with a link to the source deep-drill page.  

\section{Interface tools to communicate with robotic, follow-up telescopes}
\label{sec:robotic}

The GROWTH Marshal provides an interface between the Marshal's users and the managers of robotic telescopes for automatically requesting follow-up observations of sources saved on the Marshal, e.g., the Palomar 60-inch telescope and the Liverpool telescope. Telescope managers are able to specify what information is required to perform a follow-up observation and grant permission to some science programs for automated follow-up.  When requesting follow-up observation of a source, users who are members of approved science programs are able to complete a form generated from the telescope team's specifications which is then passed along to the telescope team.

When providing a specification of the information required for follow-up, a telescope team is required to also specify the URL to which the Marshal may send HTTP Post requests.  When a user completes a follow-up request form, the form is then sent over to this specified URL.  The GROWTH Marshal then uses the response from this Post request as an indication of whether the telescope team has accepted the request.  In the event of a successful response, the followup request is saved on the Marshal as a pending observation.  If the Post request times out or is sent to an invalid URL, the Marshal notifies the user that it was unable to communicate with the telescope team and does not save the failed follow-up request.  If the Post request returns an error code, the error description is provided to the user and the request is not saved on the Marshal.  This allows telescope teams to reject invalid or incomplete request forms and indicate the reason for rejection (e.g., if the user did not properly format some required parameter or if the telescope is completely down for maintenance).  

Once a follow-up request is submitted, the scheduling and execution of the observation is left entirely to the discretion of the telescope team.  The Marshal allows telescope teams to update the current status for submitted follow-up requests to inform the Marshal users of how close the observation is to completion.  The Marshal also allows users to modify or cancel submitted follow-up requests, relaying the changed information back to the telescope team in the same manner as used to submit the original request.  Upon the completion of a requested observation,  the telescope team is able to submit the resulting photometric or spectroscopic information to the Marshal in a standardized format.  This photometry or spectroscopy is then incorporated with the Marshal's existing saved photometric and spectroscopic data, allowing users to interact with the newly observed data with the Marshal's set of inbuilt analysis tools.

\section{Observing Run Planner for classical follow-up}
\label{sec:observingrun}

In addition to the integrated automatic follow-up of targets, the GROWTH Marshal also contains a toolset for classical follow-up. This toolset is organized as follows:

\begin{itemize}
\item The scheduler keeps track of available follow-up resources (e.g, date of observing run, telescope, instrument, observer/contact).
\item On the overview page of each target, follow-up can be assigned, with a priority (1-5), and a comment giving details of the needed follow-up.
\item For each observing date, the Observing Planner will compile a sortable list of the targets that have been assigned, and generate a starlist with offset stars for each target, for use at the telescope. 
\end{itemize}

\begin{figure*}[ht!]
\centering
\includegraphics[width=1\textwidth]{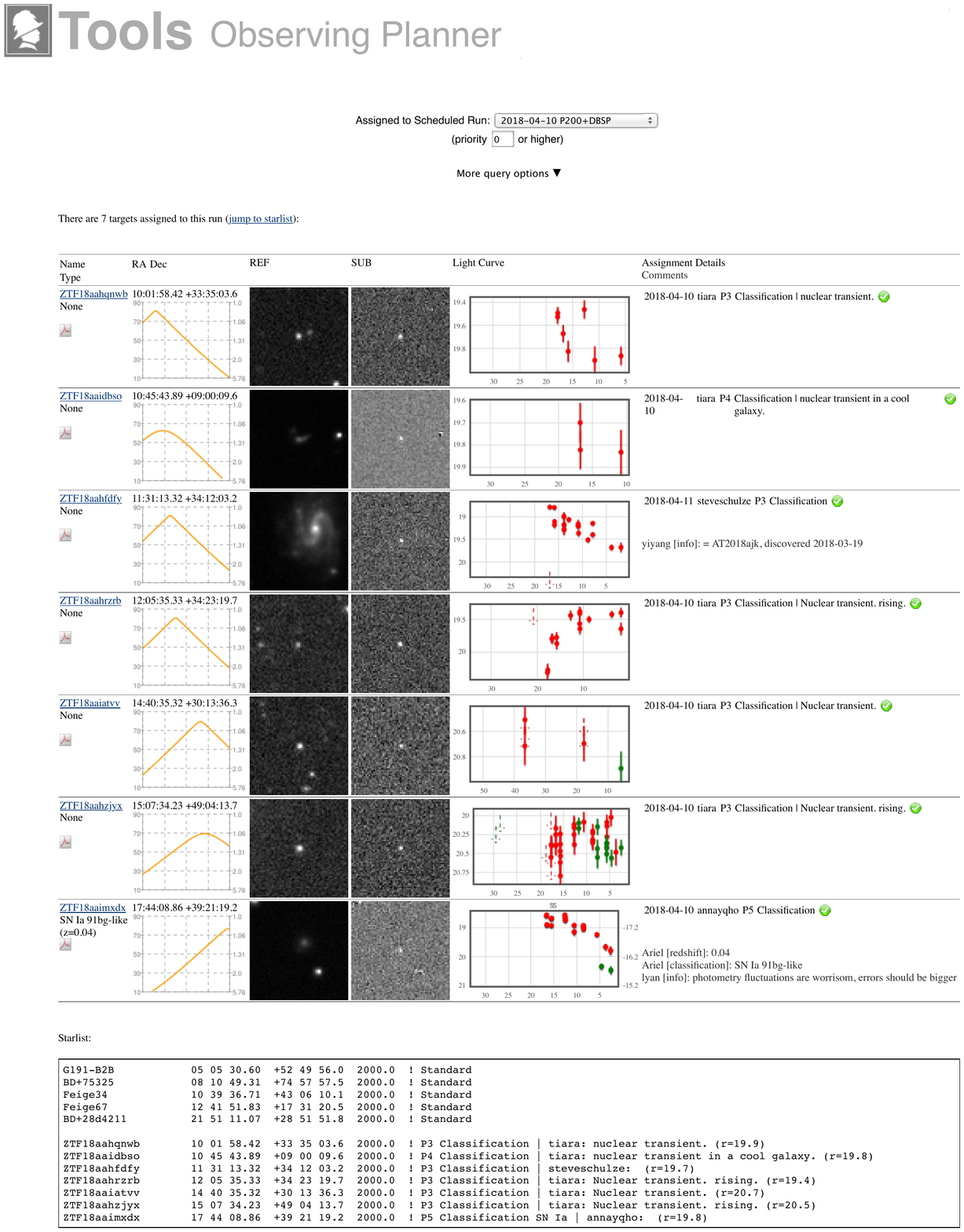}
\caption{\label{fig:op}The Observing Planner page. Green checkboxes allow the observer to mark the target as observed during the observing night. The pdf symbols in the first column generates finder charts with offset stars.}
\end{figure*}

The Observing Planner page can be used both to plan ahead and keep track in real-time of the observations that are done during the night. For each target, the Observing Planner displays the visibility plot, the lightcurve, cutouts from our reference and subtraction images, and any comments that were added when the follow-up assignments were made (see Fig.~\ref{fig:op}). Links to generate finder charts with an offset grid from three nearby, bright stars are also automatically generated.

The target list is by default sorted according to the visibility of the transient from the relevant telescope. This makes it very simple for the observer to optimize the observing schedule for the night. Sorting by RA, Dec, Name, type, age, priority, magpsf and redshift are also allowed. In case a high number of targets have been assigned to an observing run, the Observing Planner also allows for different filtering criteria, such as priority and/or candidate type (e.g, nuclear, variable star, transient) and/or classification (e.g., SN Ia, SN II, AGN) and/or science program(s). The essential information from the target list is also included in the automatically generated starlist, including the latest filtered magnitude, which allows exposure times to be estimated. Any comments added when the follow-up assignments were made are also appended to the starlist.

\section{Ingest and Visualize Follow-up Data}
\label{sec:followup}

After follow-up observations of sources in the GROWTH Marshal have been performed, the results need to be reported back to the collaboration by ingesting the observations into the Marshal database. Due to the number of follow-up facilities at the collaboration's disposal, it would be impossible to include automatic interfaces for all follow-up programs. Instead the follow-up data needs to be uploaded using two simple forms, one for photometry and another for spectroscopy. 

The photometry upload form allows the user to either submit single points of data or a whole ASCII table of observation, provided that they were all obtained with the same telescope and instrument. Data can only be submitted for combinations of telescope and instrument that are known to the Marshal database. If a telescope or instrument is missing from the database it can be added through a linked form. When submitting an ASCII table of data the user must specify which type of information each column contains. Regardless of whether a single data point or a table, certain information is always required for the upload. This includes the Julian Date of the observation, filter, magnitude, magnitude error and the limiting magnitude. Optionally the user can also list the observers and reducers of the data to allow them to be credited for their contribution.

Manually uploaded photometry is added to the photometry table of the database which also collects all the observations and limiting magnitudes of the P48 observations that were ingested from the Avro alert packets. On a dedicated photometry page of a source this data can be as an interactive plot of the lightcurve, on which the user can zoom in for close inspection, as well as a simple table that can also be exported to a comma-separated-value (csv) file for further data analysis.

The spectroscopy upload form accepts spectroscopic data as either FITS or ASCII file. The upload is not limited to spectra of the object; observations of the host galaxy or the sky can also be submitted. As for photometric data, telescope and instrument first need to be entered into the data base. If the uploaded files contain important information as observation date and exposure time in the header, this will be entered into the database automatically. If the information is missing from the header, it can be provided in the form. Again the observer and reducer can be credited there as well. If further information was derived from the spectrum such as redshift or phase of the transient, this can be included as well and will be displayed to all users as comments on the spectrum.

\begin{figure*}[ht!]
\centering
\includegraphics[width=0.8\textwidth]{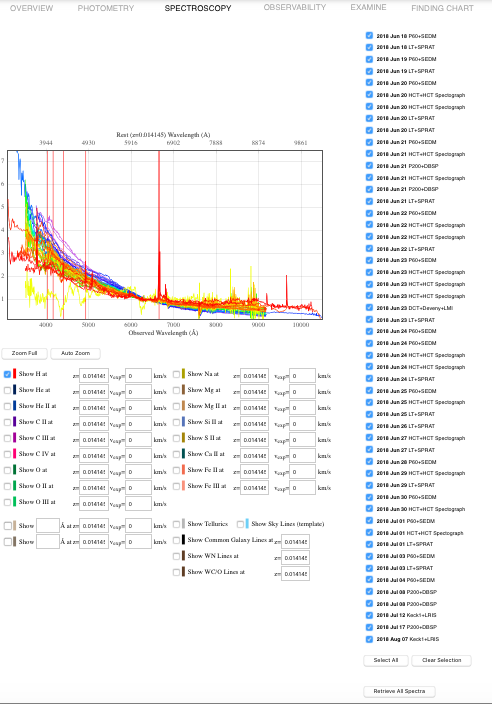}
\caption{\label{fig:spectra} Snapshot of the GROWTH marshal interactive spectra page. We allow users to enter line identifications, specify redshift and/or velocity, mark common sky/telluric/galaxy lines, select/unselect a subset of the spectroscopic sequence.}
\end{figure*}

The uploaded spectroscopic data can be viewed on a dedicated spectroscopy page for the source (Figure~\ref{fig:spectra}). There the spectra can be inspected in interactive plots of either multiple spectra or the individual ones. Like the photometric lightcurve plots, the user can can zoom in to relevant parts of the spectrum. The most relevant absorption lines (as well as common galactic and atmospheric lines) can be overlayed and shifted to the transient redshift and expansion velocity. Furthermore the individual spectra can be binned into larger bins to make the inspection of noisy data easier.

\section{View Reports of Properties of Selected Data}
\label{sec:report}

\begin{figure*}[ht!]
\centering
\includegraphics[width=0.8\textwidth]{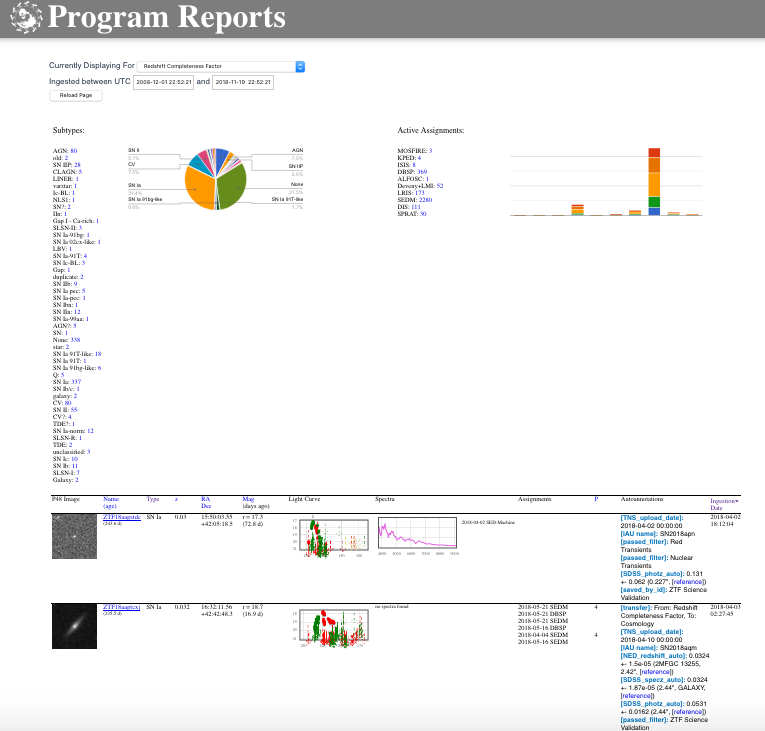}
\caption{\label{fig:reports} Snapshot of the GROWTH Marshal reports page. The pie-chart shows spectroscopic classifications and allows filtering by type. The histogram shows follow-up assignments to various follow-up telescopes. The table summarizes various properties of the transients and can be exported into machine-readable format.}
\end{figure*}

Summary of all sources belonging to a user's science programs can be viewed on the ``Program Reports'' page, on which the user can either view all of their science programs or be filtered down to a single program (Figure~\ref{fig:reports}). This page contains a table of the basic data of the transient, including type, redshift, RA, Dec and the latest observed magnitude. Additionally small previews of the lightcurves and spectroscopic data (if available) are provided for each transient. Lastly there are columns for current follow-up assignments and their priority, autoannotations and ingestion date that have been added to the source. The table can be sorted by most columns and additionally by any autoannotation by clicking on the autoannotation name. The sorting by autoannotation allows each science program to define their own metric by which the candidates are to be sorted. Sources can also be filtered based on ingestion date range and/or follow-up instruments and/or classification. For further management of a science program, a pie-chart shows the classification of all the sources assigned to the program.

\section{Triggering a Target of Opportunity for Multi-messenger Astrophysics}
\label{sec:too}

The Target of Opportunity (ToO) Marshal is designed to facilitate the multi-messenger detection of counterparts to short gamma-ray burst, high energy neutrino, and gravitational-wave candidate events. The workflow of the ToO Marshal begins with a trigger from an external source outside of the GROWTH marshal. An external instrument, such as the Fermi Gamma-Ray Burst Monitor (GBM), IceCube Neutrino Observatory, or Advanced LIGO and Advanced Virgo, releases a candidate event by Gamma-ray Coordinates Network (GCN). These GCN notices include information about the candidate, including sky localization information, either in the form of a Hierarchical Equal Area isoLatitude Pixelization (HEALPix) map or center of sky localization and error ellipse, and other identifying features, such as whether the gamma-ray burst is short or long, or the likelihood that the gravitational-wave transient has an object with a mass consistent with a neutron star. For any given transient, which are uniquely identified by their time, a series of ``tags'' are created, which give short, characteristic representations of the transients. For example, Fermi GBM transients have tags such as Fermi, short, long, and GRB, which are applied depending on the information in the notice.

For each of these notices, pages are created with five overall sections. The first, which applies in the Fermi GBM case, is to show a plot of the lightcurve. The second is to demonstrate the observability of the source. This includes the time windows that the source is available from Palomar and the time elapsed since the trigger time. The third are the GCN notices associated with the transient. For many of these transients, refined analyses lead to improvements in the sky localization. For example, in the gravitational-wave case, incorporation of detector calibration errors and full MCMC-based analyses can change the localization. For this reason, each notice is ingested and analyzed separately. The fourth is a section for observing plans, which are created for each of these notices. Finally, links to the GROWTH Marshal are created corresponding to the particular observing program.

There are a few components to each observing plan. The first is a queue name, which identifies the name given to the particular set of observations. The second is the start and end time of the set of observations. The third is the maximum exposure time for the observations. The fourth is the filter the observations will be taken in. A schedule with the fields encompassing the probability region are created based on these requested parameters. The scheduling is optimized using algorithms described in \citealt{Rana2017, Ghosh2017}. Subsequent sets of exposures of the same fields (potentially in different filters) are performed by making a separate request outside of the original window of time requested.

\section{Conclusion}
In summary, the GROWTH marshal provides an effective collaborative platform for time-domain astronomy. This portal continues to be developed based on feedback from the scientists using it with new features added every week. The GROWTH Marshal facilitates full automation of alert saving and alert follow-up. If a science program decides their alert stream is sufficiently pure that they would like to fully automate saving alerts and fully automate follow-up, this is technically straightforward. Indeed, two science programs are currently exploring full automation.

As we prepare for the LSST era, where the alert stream increases by yet another order of magnitude, science portals like the GROWTH Marshal will serve to assist collaborations focus on answering science questions and prioritize use of follow-up resources amidst the deluge of transient alerts.  

\bigskip
\bigskip
This work was supported by the GROWTH (Global Relay of Observatories Watching Transients Happen) project funded by the National Science Foundation Partnership in International Research and Education program under Grant No 1545949. GROWTH is a collaborative project between California Institute of Technology (USA), Pomona College (USA), San Diego State University (USA), Los Alamos National Laboratory (USA), University of Maryland College Park (USA), University of Wisconsin Milwaukee (USA), Tokyo Institute of Technology (Japan), National Central University (Taiwan), Indian Institute of Astrophysics (India), Inter-University Center for Astronomy and Astrophysics (India), Weizmann Institute of Science (Israel), The Oskar Klein Centre at Stockholm University (Sweden), Humboldt University (Germany). Tested with observations obtained with the Samuel Oschin Telescope 48-inch and the 60-inch Telescope at the Palomar Observatory as part of the Zwicky Transient Facility project. Major funding has been provided by the U.S National Science Foundation under Grant No. AST-1440341 and by the ZTF partner institutions: the California Institute of Technology, the Oskar Klein Centre, the Weizmann Institute of Science, the University of Maryland, the University of Washington, Deutsches Elektronen-Synchrotron, the University of Wisconsin-Milwaukee, and the TANGO Program of the University System of Taiwan. We thank I. Arcavi for valuable contributions to the predecessor of this system, the Palomar Transient Factory marshal system. 


\end{document}